\def\etal{{\frenchspacing\it et al.}}

\def\beq#1{\begin{equation}\label{#1}}
\def\eeq{\end{equation}}
\def\beqa#1{\begin{eqnarray}\label{#1}}
\def\eeqa{\end{eqnarray}}

\def\fun#1#2{\lower3.6pt\vbox{\baselineskip0pt\lineskip.9pt
        \ialign{$\mathsurround=0pt#1\hfill##\hfil$\crcr#2\crcr\sim\crcr}}}
        
\def\xi{{{\bf x}^b}}

\def\rmdet{{\rm det}}
\def\rmCov{{\rm Cov}}
\def\rmFoM{{\rm FoM}}
\def\zbar{\overline{z}}

\documentclass[twocolumn,aps,showpacs,showkeys,nofootinbib]{revtex4}
\usepackage{epsfig}

\newcommand{\be}{\begin{equation}}
\newcommand{\ee}{\end{equation}}
\newcommand{\ba}{\begin{eqnarray}}
\newcommand{\ea}{\end{eqnarray}}

\begin{document}
\input{epsf.sty}

\title{Figure of Merit for Dark Energy Constraints
from Current Observational Data}
\author{Yun~Wang}
\address{Homer L. Dodge Department of Physics \& Astronomy, Univ. of Oklahoma,
                 440 W Brooks St., Norman, OK 73019;
                 email: wang@nhn.ou.edu}

                 \today

\begin{abstract}
In order to make useful comparisons of different dark energy experiments,
it is important to choose the appropriate figure of merit (FoM) 
for dark energy constraints.
Here we show that for a set of dark energy parameters $\{f_i\}$,
it is most intuitive to define FoM=$1/\sqrt{\rmdet\, \rmCov(f_1,f_2,f_3,...)}$,
where $\rmCov(f_1,f_2,f_3,...)$ is the covariance matrix of $\{f_i\}$.
In order for this FoM to represent the dark energy constraints 
in an optimal manner, the dark energy parameters $\{f_i\}$
should have clear physical meaning, and be minimally correlated.
We demonstrate two useful choices of $\{f_i\}$
using 182 SNe Ia (from the HST/GOODS program, 
the first year Supernova Legacy Survey, and nearby SN Ia surveys), 
[$R(z_*)$, $l_a(z_*)$, $\Omega_b h^2$] from the five year 
Wilkinson Microwave Anisotropy Probe (WMAP) observations,
and SDSS measurement of the baryon acoustic oscillation (BAO) scale,
assuming the HST prior of $H_0=72\pm 8\,$(km/s)Mpc$^{-1}$,
and without assuming spatial flatness.
We find that for a dark energy equation of state linear in the
cosmic scale factor $a$, the correlation of $(w_0,w_{0.5})$ 
[$w_0=w_X(z=0)$, $w_{0.5}=w_X(z=0.5)$, with
$w_X(a)=3 w_{0.5}-2 w_0 + 3( w_0 - w_{0.5} ) \,a$] is 
significantly smaller than that of $(w_0,w_a)$ [with 
$w_X(a)=w_0 + (1-a) w_a$].
In order to obtain model-independent constraints on dark energy,
we parametrize the dark energy density function
$X(z)=\rho_X(z)/\rho_X(0)$ as a free function
with $X_{0.5}$, $X_{1.0}$, and $X_{1.5}$ [values of $X(z)$
at $z=0.5$, 1.0, and 1.5] as free parameters estimated from data.
If one assumes a linear dark energy equation of state, 
current observational data are consistent 
with a cosmological constant at 68\% C.L.
If one assumes $X(z)$ to be a free function parametrized by 
($X_{0.5}$, $X_{1.0}$, $X_{1.5}$), current data 
deviate from a cosmological constant at $z=1$ at 68\% C.L.,
but are consistent with a cosmological constant at 95\% C.L..
Future dark energy experiments will allow us to dramatically
increase the FoM of constraints on ($w_0,w_{0.5})$, and of ($X_{0.5}, X_{1.0}, X_{1.5}$).
This will significantly shrink the dark energy parameter space
to either enable the discovery of dark energy evolution,
or the conclusive evidence for a cosmological constant.

\end{abstract}



\maketitle


\section{Introduction}

The understanding of dark energy, the unknown cause for the observed 
cosmic acceleration \cite{Riess98,Perl99}, continues to be 
one of the most important challenges in cosmology today.
Dark energy could be an unknown energy component
\cite{Freese87,Linde87,Peebles88,Wett88,Frieman95,Caldwell98},
or a modification of general relativity 
\citep{SH98,Parker99,Boisseau00,DGP00,Freese02}.
\cite{Pad} and \cite{Peebles03} contain reviews of many models.
Much work continues to be done on the theoretical front,
see for example, \cite{OW04,Cardone05,Caldwell06,KO06,DeFelice07,Koi07}.
Current observational data do not provide stringent constraints
on dark energy, and allow a wide range of possibilities including 
dark energy being a cosmological constant (see, for example,
\cite{Wang04,WangTegmark04,WangTegmark05,Alam05,Daly05,Jassal05b,Polar05,Barger06,Dick06,Huterer06,Ichi06,Jassal06,Liddle06,Nesseris06,Schimd06,Sumu06,WangPia06,Wilson06,Xia06,Alam07,Clarkson07,Davis07,Gong07,Kazu07,Wei07,Wright07,Zhang07,Zun07}).

Future dark energy experiments that are significantly more ambitious
than current ones are required to illuminate the nature of
dark energy. In order to compare proposed future dark energy
experiments in a useful manner, we need to choose the appropriate 
figure of merit (FoM) for dark energy constraints \cite{FoM}.

In this paper, we explore the optimization of FoM using current
observational data from supernovae, galaxy clustering,
and cosmic microwave background anisotropy (CMB) data.
We describe our method in Sec.II, present our results in Sec.III,
and conclude in Sec.IV.

\section{Method}

\subsection{General definition for figure of merit}

When we estimate a set of parameters, $\{f_i\}$ (i=1, 2, ..., $N$), from data, 
the most intuitive figure of merit is the $N$-dimensional volume
enclosed by the 68\% or 95\% confidence level (C.L.) contours of the parameters.
If the likelihood surfaces for all the parameters are Gaussian,
this $N$-dimensional volume is proportional to the squre root
of the covariance matrix of $\{f_i\}$, $\sqrt{ \rmdet\,\rmCov(f_1,f_2,f_3,...)}$.
For $N=2$, the 68\% or 95\% C.L. contours of $f_1$ and $f_2$ 
are ellipses, with enclosed area given by $\pi \sqrt{ \rmdet\,\rmCov(f_1,f_2)}$
multiplied by 2.30 or 6.17. Parametrizing dark energy equation of state as 
$w_X(a)=w_0+(1-a)w_a$ \cite{Chev01}, the Dark Energy Task Force (DETF) 
defined FoM to be the inverse of the area enclosed by the 95\% C.L. 
contour of ($w_0,w_a$) \cite{detf}, i.e., 
\be
\rmFoM_{DETF}=\frac{1}{6.17 \pi \sigma({w_a}) \sigma({w_p})}
\ee
where $w_p=w_0- w_a\, \langle \delta w_0 \delta w_a\rangle/\langle \delta w_a^2\rangle$,
and $\sigma({w_i})=\sqrt{\langle \delta w_i^2\rangle}$. 
Note that $\sigma({w_a}) \sigma({w_p})=\sqrt{ \rmdet\,\rmCov(w_0,w_a)}$,
thus the conversion to $w_p$ is $\it not$ needed to calculate the FoM.

For real data, the likelihood surfaces for the parameters $\{f_i\}$ are almost 
always {\it non-Gaussian} at the 95\% C.L., thus defining the FoM 
as enclosed area or volume by the 95\% C.L. contours 
of $\{f_i\}$ becomes problematic.
We propose the definition for a relative generalized FoM
given by
\be
\rmFoM_r=\frac{1}{\sqrt{ \rmdet\,\rmCov(f_1,f_2,f_3,...)}},
\label{eq:FoM_r}
\ee
where $\{f_i\}$ are the chosen set of dark energy parameters.
This definition has the advantage of being easy to calculate for
either real or simulated data. We have streamlined the definition
to omit numerical factors since what matters is the relative FoM between
different experiments.

Note that while this FoM definition has an intuitive physical
interpretation, it rewards experiments that yield very correlated
estimates of the dark energy parameters. This is especially
true in applying the DETF FoM, since ($w_0, w_a$) are always
highly correlated. Hence the dark energy FoM [as defined
in Eq.(\ref{eq:FoM_r})] is most meaningful
when the dark energy parameters $\{f_i\}$ are chosen such that 
they are minimally correlated with each other.


\subsection{Dark energy parametrization}

We study constraints on a 2-parameter dark energy
equation of state $w_X(z)$ linear in $a$, as well 
the dark energy density function
$X(z) \equiv\rho_X(z)/\rho_X(0)$
as a free function at $z\leq 1.5$.

The 2-parameter $w_X(z)$ is given by
\ba
w_X(a)&=&\left(\frac{a_c-a}{a_c-1}\right) w_0
+\left(\frac{a-1}{a_c-1}\right) w_c \nonumber\\
&=&\frac{a_cw_0-w_c + a(w_c-w_0)}{a_c-1}
\label{eq:wc}
\ea
where $w_0=w_X(z=0)$, and $w_c=w_X(z=z_c)$.
Eq.(\ref{eq:wc}) corresponds to a dark energy density function
\ba
X(z) &=&\exp \left\{ 3\left[ 1+\left(\frac{a_cw_0-w_c}{a_c-1}\right)\right]\,\ln(1+z)
+ \right.\nonumber\\
& & \hskip 1cm \left.3 \left(\frac{w_c-w_0}{a_c-1}\right) \frac{z}{1+z}\right\}
\ea

Eq.(\ref{eq:wc}) is related to $w_X(z)=w_0+(1-a)w_a$ by setting
\be
w_a= \frac{w_c-w_0}{1-a_c}, \hskip 1cm
{\rm or} \hskip 1cm
w_c=w_0+(1-a_c) w_a.
\label{eq:wa,wc}
\ee
If we choose $a_c=1+\sigma^2(w_0)/\sigma^2(w_0w_a)$,
then $(w_0,w_c)$ are {\it uncorrelated}.
For current data, $z_c \sim 0.3$.
Choosing $a_c$ to make $(w_0,w_c)$ uncorrelated has
the disadvantage that $a_c$ is different for different
data sets.

We recommend choosing $z_c=0.5$; it is sufficiently close
to $z_c \sim 0.3$ that the correlation of $w_0$ and $w_{0.5}=w_X(z=0.5)$
is relatively small.
It is straightforward to show that if 
$|\sigma^2(w_0 w_a)/[\sigma(w_0)\sigma(w_a)]|<1$,
$(w_0,w_c)$ are {\it less} correlated than $(w_0,w_a)$ if
\be
\sigma^2(w_0) < 2 \left|(1-a_c) \sigma^2(w_0 w_a)\right|.
\ee
This is always satisfied for $z_c=0.5$.
Choosing $z_c=0.5$, the correlation
of $(w_0,w_{0.5})$ is smaller than that of $(w_0,w_a)$
by about a factor of 2 for the combined SNe, BAO, and CMB data
considered in this paper.
Fixing $z_c$ has the significant advantage of allowing the
comparison of the {\it same} dark energy property 
for different data sets. 
For our results for the 2-parameter dark energy model,
we use Eq.(\ref{eq:wc}) with $a_c=1/(1+0.5)=2/3$ ($z_c=0.5$).
Thus
\ba
\label{eq:w_0.5}
&&w_X(a)=3 w_{0.5}-2 w_0 + 3\left( w_0 - w_{0.5} \right) \,a\\
&&X(z)= (1+z)^{3(1-2 w_0+ 3w_{0.5})} \exp\left[9(w_0-w_{0.5})\, \frac{z}{1+z}\right]
\nonumber
\ea

In order to obtain model-independent constraints on dark energy,
we parametrize the dark energy density function
$X(z)=\rho_X(z)/\rho_X(0)$ as a free functions
with $X_{0.5}$, $X_{1.0}$, and $X_{1.5}$ [values of $X(z)$
at $z=0.5$, 1.0, and 1.5] as free parameters estimated from data.
At $z>1.5$, we choose either $X(z)=X_{1.5}$, or
$X(z)=X_{1.5} \,e^{\alpha (z-1.5)}$ (with $\alpha$ as an
additional parameter to be estimated from data). Our results are insensitive
to the assumption about $X(z)$ at $z>1.5$ (other than that
dark energy becomes insignificant at $z>1.5$).
As more data become available at $z>1.5$, we can include
$X_{2.0}=X(z=2.0)$, $X_{2.5}=X(z=2.5)$, and
$X_{3.0}=X(z=3.0)$ as estimated parameters,
as well as inserting more estimated $X(z)$ values at
$z<1.5$.
Early dark energy (significant at high $z$) is not required by current data,
and leads to contradiction with observed cosmic structure formation \cite{Sandvik},
unless a cutoff is imposed.

The constraints on $X_{0.5}$, $X_{1.0}$, and $X_{1.5}$ are
insensitive to the interpolation used in deriving $X(z)$ elsewhere.
The simplest smooth interpolation is given by a polynomial:
\ba
X(z)&=& -\frac{1}{2} \left(3\zbar-1\right) \left(3\zbar-2\right) 
\left(\zbar-1\right)\nonumber\\
& & \hskip 0.2cm +\frac{9}{2} X_{0.5}\zbar \left(3\zbar-2\right) 
\left(\zbar-1\right) \nonumber\\
& & \hskip 0.2cm -\frac{9}{2} X_{1.0} \zbar \left(3\zbar-1\right) 
\left(\zbar-1\right)\nonumber\\
& & \hskip 0.2cm +\frac{1}{2} X_{1.5} \zbar \left(3\zbar-1\right) \left(3\zbar-2\right),
\label{eq:Xz}
\ea
where $\zbar=z/1.5$. 

\subsection{Data analysis technique}

The comoving distance from the observer to redshift $z$ is given by
\ba
\label{eq:rz}
&&r(z)=cH_0^{-1}\, |\Omega_k|^{-1/2} {\rm sinn}[|\Omega_k|^{1/2}\, \Gamma(z)],\\
&&\Gamma(z)=\int_0^z\frac{dz'}{E(z')}, \hskip 1cm E(z)=H(z)/H_0 \nonumber
\ea
where $\Omega_k=-k/H_0^2$ with $k$ denoting the curvature constant, 
and ${\rm sinn}(x)=\sin(x)$, $x$, $\sinh(x)$ for 
$\Omega_k<0$, $\Omega_k=0$, and $\Omega_k>0$ respectively, and
\be
E^2(z)=\Omega_m (1+z)^3+\Omega_{\rm rad}(1+z)^4+\Omega_k(1+z)^2+
\Omega_X X(z)
\ee
with $\Omega_X=1-\Omega_m-\Omega_{\rm rad}-\Omega_k$, and the dark energy density
function $X(z) \equiv \rho_X(z)/\rho_X(0)$.

CMB data give us the comoving distance to the photon-decoupling surface 
$r(z_*)$, and the comoving sound horizon 
at photo-decoupling epoch \cite{EisenHu98,Page03}
\ba
\label{eq:rs}
r_s(z_*) &=& \int_0^{t_*} \frac{c_s\, dt}{a}
=cH_0^{-1}\int_{z_*}^{\infty} dz\,
\frac{c_s}{E(z)}, \nonumber\\
&=& cH_0^{-1} \int_0^{a_{*}} 
\frac{da}{\sqrt{ 3(1+ \overline{R_b}\,a)\, a^4 E^2(z)}},
\ea
where $a$ is the cosmic scale factor, 
$a_* =1/(1+z_*)$, and
$a^4 E^2(z)=\Omega_m (a+a_{\rm eq})+\Omega_k a^2 +\Omega_X X(z) a^4$,
with $a_{\rm eq}=\Omega_{\rm rad}/\Omega_m=1/(1+z_{\rm eq})$, and
$z_{\rm eq}=2.5\times 10^4 \Omega_m h^2 (T_{CMB}/2.7\,{\rm K})^{-4}$.
The sound speed is $c_s=1/\sqrt{3(1+\overline{R_b}\,a)}$,
with $\overline{R_b}\,a=3\rho_b/(4\rho_\gamma)$,
$\overline{R_b}=31500\Omega_bh^2(T_{CMB}/2.7\,{\rm K})^{-4}$.
COBE four year data give $T_{CMB}=2.728\pm 0.004\,$K (95\%
C.L.) \cite{Fixsen96}.
We take $T_{CMB}=2.725$ following \cite{Komatsu08}, since
we will use the CMB bounds derived by \cite{Komatsu08}.
The angular scale of the sound horizon at recombination is
defined as $l_a=\pi r(z_*)/r_s(z_*)$ \cite{Page03}.

Wang \& Mukherjee 2007 \cite{WangPia07} showed that
the CMB shift parameters
\be
R \equiv \sqrt{\Omega_m H_0^2} \,r(z_*), \hskip 0.1in
l_a \equiv \pi r(z_*)/r_s(z_*),
\ee
together with $\Omega_b h^2$, provide an efficient summary
of CMB data as far as dark energy constraints go.
We use the covariance matrix of [$R(z_*), l_a(z_*), \Omega_b h^2]$ from
the five year WMAP data (Table 11 of \cite{Komatsu08}), with $z_*$
given by fitting formulae from Hu \& Sugiyama (1996) \cite{Hu96}:
\be
z_*=1048\, \left[1+ 0.00124 (\Omega_b h^2)^{-0.738}\right]\,
\left[1+g_1 (\Omega_m h^2)^{g_2} \right],
\ee
where
\ba
g_1&=&\frac{0.0783\, (\Omega_b h^2)^{-0.238}}
{1+39.5\, (\Omega_b h^2)^{0.763}}\\
g_2&=&\frac{0.560}{1+21.1\, (\Omega_b h^2)^{1.81}}
\ea
CMB data are included in our analysis by adding
the following term to the $\chi^2$ of a given model
with $p_1=R(z_*)$, $p_2=l_a(z_*)$, and $p_3=\Omega_b h^2$:
\be
\label{eq:chi2CMB}
\chi^2_{CMB}=\Delta p_i \left[ Cov^{-1}(p_i,p_j)\right]
\Delta p_j,
\hskip .5cm
\Delta p_i= p_i - p_i^{data},
\ee
where $p_i^{data}$ are the maximum likelyhood values given in 
Table 10 of \cite{Komatsu08}.

SN Ia data give the luminosity distance as a function of redshift,
$d_L(z)=(1+z)\, r(z)$.
We use 182 SNe Ia from the HST/GOODS program \cite{Riess07} and the first 
year SNLS \cite{Astier05}, together with nearby SN Ia data,
as compiled by \cite{Riess07}.
We do not include the ESSENCE data \cite{Wood07}, as these are not yet derived using
the same method as thosed used in \cite{Riess07}.
Combining SN Ia data derived using different analysis techniques 
leads to systematic effects in the estimated SN distance moduli
\cite{Wang00b,Wood07}.
Appendix A of \cite{WangPia07} describes in detail how we use 
SN Ia data (flux-averaged to reduce lensing-like systematic 
effects \cite{Wang00b,WangPia04,Wang05}
and marginalized over $H_0$) in this paper.

We also use the SDSS baryon acoustic oscillation (BAO)
scale measurement by adding the following term to the
$\chi^2$ of a model:
\be
\chi^2_{BAO}=\left[\frac{(A-A_{BAO})}{\sigma_A}\right]^2,
\label{eq:chi2bao}
\ee
where $A$ is defined as
\be
\label{eq:A}
A = \left[ r^2(z_{BAO})\, \frac{cz_{BAO}}{H(z_{BAO})} \right]^{1/3} \, 
\frac{\left(\Omega_m H_0^2\right)^{1/2}} {cz_{BAO} },
\ee
and $A_{BAO}=0.469\,(n_S/0.98)^{-0.35}$,
$\sigma_A= 0.017$, and $z_{BAO}=0.35$
(independent of a dark energy model) \cite{Eisen05}. 
We take the scalar spectral index $n_S=0.96$ as measured by WMAP
five year observations \cite{Komatsu08}.

For Gaussian distributed measurements, the likelihood function
$L\propto e^{-\chi^2/2}$, with 
\be
\chi^2=\chi^2_{CMB}+\chi^2_{SNe}+\chi^2_{BAO},
\label{eq:chi2}
\ee
where $\chi^2_{CMB}$ is given in Eq.({\ref{eq:chi2CMB}}),
$\chi^2_{SNe}$ is given in Appendix A
of \cite{WangPia07}, and $\chi^2_{BAO}$ is given in Eq.({\ref{eq:chi2bao}}).

The current BBN constraints \cite{Steigman06}, $S=0.942\pm 0.030$
($N_{\nu}=2.30^{+0.35}_{-0.34}$) rule out the standard model 
of particle physics ($S=1$, $N_{\nu}=3$) at 1$\sigma$ \cite{Steigman06}.
Given the uncertainties involved in deriving the BBN constraints,
we relax the standard deviation of $S$ by a factor of two, so that the 
standard model of particle physics is allowed at 1$\sigma$.
We find that the resultant BBN constraints do not have measurable effect 
on our dark energy constraints.

For all the dark energy constraints from combining the different
data sets presented in this paper, 
we marginalize the SN Ia data over $H_0$ in
flux-averaging statistics \cite{WangPia07}, and 
impose a prior of $H_0=72\pm 8\,$(km/s)Mpc$^{-1}$
from the HST Cepheid variable star observations \cite{HST_H0}.

We run a Monte Carlo Markov Chain (MCMC) based on the MCMC engine 
of \cite{Lewis02} to obtain ${\cal O}$($10^6$) samples for each set of 
results presented in this paper. 
The parameters used are ($\Omega_k$, $\Omega_m$, $h$, $\Omega_b h^2$, 
$\mbox{\bf p}_{DE}$). The dark energy parameter set 
is described in Sec.IIB.
We assumed flat priors for all the parameters, and allowed ranges 
of the parameters wide enough such that further increasing the allowed 
ranges has no impact on the results.
The chains typically have worst e-values (the
variance(mean)/mean(variance) of 1/2 chains)
much smaller than 0.005, indicating convergence.
The chains are subsequently 
appropriately thinned to ensure independent samples.

\section{Results}

\begin{figure} 
\psfig{file=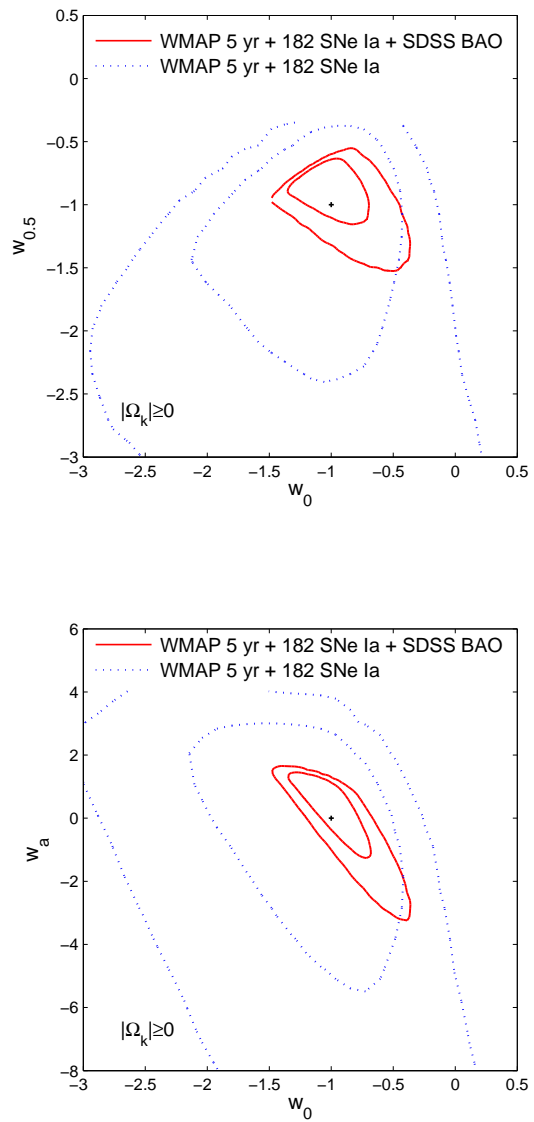,width=2.8in}\\
\caption[2]{\label{fig:w0wc}\footnotesize%
The 68\% and 95\% C.L. contours of
($w_0,w_{0.5}$) (upper panel) and ($w_0,w_a$) (lower panel) from
WMAP 5 year measurement of [$R(z_*)$, $l_a(z_*)$, $\Omega_b h^2$],
and 182 SNe Ia (from the HST/GOODS program, 
the first year Supernova Legacy Survey, and nearby SN Ia surveys), 
with and without the SDSS measurement of the baryon acoustic oscillation 
(BAO) scale.}
\end{figure}

Fig.{\ref{fig:w0wc}} shows the 68\% and 95\% C.L. contours of
($w_0,w_{0.5}$) (upper panel) and ($w_0,w_a$) (lower panel) from
WMAP 5 year measurement of [$R(z_*)$, $l_a(z_*)$, $\Omega_b h^2$],
and 182 SNe Ia (from the HST/GOODS program, 
the first year Supernova Legacy Survey, and nearby SN Ia surveys), 
with and without the SDSS measurement of the baryon acoustic oscillation 
(BAO) scale.
We have assumed the HST prior of $H_0=72\pm 8\,$(km/s)Mpc$^{-1}$,
and allowed $\Omega_k$ to vary.
Table 1 shows the mean, rms variance, and correlation coefficients 
of ($w_0,w_{0.5}$) and ($w_0,w_a$), as well as the relative 
dark energy FoM$_r$ defined in Eq.(\ref{eq:FoM_r}). 
Note that Pearson's correlation coefficient
$\rho_{xy}=\sigma^2(x y)/[\sigma(x)\sigma(y)]$.
Adding the SDSS BAO scale measurement improves
the FoM$_r$ by a factor of 21.5 for ($w_0,w_{0.5}$), and
by a factor of 27.0 for ($w_0,w_a$). Since ($w_0,w_{0.5}$)
are significantly less correlated than ($w_0,w_a$),
the improvement factor in FoM$_r$ of ($w_0,w_{0.5}$) is a more
reliable indicator of the impact of adding the SDSS BAO scale measurement.

\begin{table*}[htb]
\caption{Constraints on ($w_0,w_{0.5}$) and ($w_0,w_a$)}
\begin{center}
\begin{tabular}{llllllll}
\hline
Data& $\mu(w_0)$ & $\sigma(w_0)$ & $\mu(w_0)$ & $\sigma(w_{0.5})$ & $\rho_{w_0w_{0.5}}$  & FoM$_r$ \\
WMAP5+SNe & -1.075 & 0.598 & -1.939 & 1.572 & -0.401 & 1.163 \\
WMAP5+SNe+BAO & -0.937 & 0.226 & -0.953 & 0.206 & -0.512 & 25.013 \\
\hline
 \hline	
Data& $\mu(w_0)$ & $\sigma(w_0)$ & $\mu(w_a)$ & $\sigma(w_a)$ & $\rho_{w_0w_a}$  & FoM$_r$ \\
WMAP5+SNe & -1.073 &  0.647 & -2.960 & 6.759 & -0.670 & 0.308 \\
WMAP5+SNe+BAO & -0.938 & 0.226 & -0.045 & 1.126 & -0.882 & 8.326 \\
\hline
\end{tabular}
\end{center}
\end{table*}

Fig.2 shows the one dimensional marginalized probability distributions
(pdf) of ($\Omega_m$, $h$, $\Omega_k$, $\Omega_b h^2$, $w_0$, $w_a$),
for 182 SNe Ia, the SDSS BAO scale measurement,
and the WMAP 5 year data in the form of measured
(1) [$R(z_*)$, $l_a(z_*)$, $\Omega_b h^2$] (solid lines),
(2) [$R(z_*)$, $l_a(z_*)$, $\Omega_b h^2$] with $z_*$ fixed
at 1090.4 (dotted),
and (3) [$R(z_*)$, $l_a(z_*)$, $z_*$] (dashed).
For reference, the dot-dashed line shows the pdfs
for 182 SNe Ia, the SDSS measurement of the BAO scale, and the 
WMAP 3 year data in the form of measured
[$R(z_{CMB})$, $l_a(z_{CMB})$, $\Omega_b h^2$] with $z_{CMB}$ fixed
at 1089 from \cite{WangPia07}.

\begin{figure} 
\psfig{file=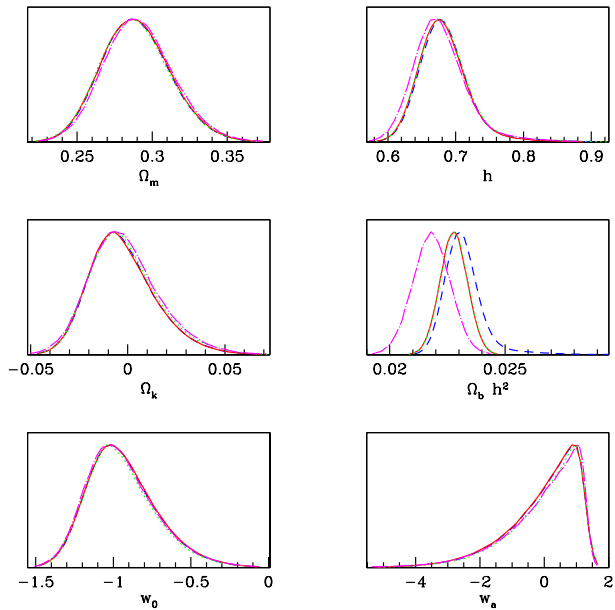,width=3.5in}\\
\caption[2]{\label{fig:w0wa_pdf}\footnotesize%
One dimensional marginalized pdfs of 
($\Omega_m$, $h$, $\Omega_k$, $\Omega_b h^2$, $w_0$, $w_a$)
from 182 SNe Ia, the SDSS BAO scale measurement, and  
the WMAP 5 year data in the form of measured
(1) [$R(z_*)$, $l_a(z_*)$, $\Omega_b h^2$] (solid lines),
(2) [$R(z_*)$, $l_a(z_*)$, $\Omega_b h^2$] with $z_*$ fixed
at 1090.4 (dotted),
and (3) [$R(z_*)$, $l_a(z_*)$, $z_*$] (dashed).
The dot-dashed line shows the pdfs
for 182 SNe Ia, the SDSS measurement of the BAO scale, and the 
WMAP 3 year data in the form of measured
[$R(z_{CMB})$, $l_a(z_{CMB})$, $\Omega_b h^2$] with $z_{CMB}$ fixed
at 1089 from \cite{WangPia07}.}
\end{figure}

We find that in spite of the different pdfs for $\Omega_b h^2$,
using the [$R(z_*)$, $l_a(z_*)$, $\Omega_b h^2$] and
[$R(z_*)$, $l_a(z_*)$, $z_*$] measurements give about the same constraints
on ($\Omega_m$, $h$, $\Omega_k$, $w_0$, $w_a$).
Using the [$R(z_*)$, $l_a(z_*)$, $\Omega_b h^2$] measurement
with $z_*$ fixed at 1090.4 gives slightly tighter constraints
on ($w_0$, $w_a$).
In combination with the supernova and BAO data,
the WMAP 5 year data improve constraints on ($w_0$, $w_a$) slightly
compared to the WMAP 3 year data, while tightening the constraints
on $\Omega_k$ and $h$.

Fig.3 shows the constraints on the dark energy density function
$X(z)=\rho_X(z)/\rho_X(0)$ parametrized by Eq.(\ref{eq:Xz}),
with $X(z)$ at $z>1.5$ given by either $X(z)=X_{1.5}$
or $X(z)=X_{1.5}\,\exp[\alpha(z-1.5)]$.
Note that the assumption about dark energy at $z>1.5$ has
only a weak effect on the dark energy constraints at $z\leq 1.5$.
Note that taking $X(z>1.5)=X_{1.5}$ gives slightly less stringent constraints
on dark energy at $z\leq 1.5$. 
This is because parametrizing dark energy at $z>1.5$ with an extra parameter 
requires choosing the early dark energy parametrization such that
it is not degenerate with cosmic curvature; this is why
$\Omega_k$ is not well constrained if we choose 
$X(z>1.5)=X_{1.5}\,(1+z)^{\alpha}$, but $\Omega_k$ is well constrained
if we choose $X(z>1.5)=X_{1.5}\,\exp[\alpha(z-1.5)]$ \cite{WangPia07}.
The latter helps break the degeneracy of $\Omega_k$ with $X(z)$,
thus leading to much tighter constraints on $\Omega_k$
and slightly tighter constraints on $X(z)$ at $z\leq 1.5$ (see Fig.3).
This suggests that the more conservative approach in constraining 
dark energy is to assume that $X(z>1.5)=X_{1.5}$.
\begin{figure} 
\psfig{file=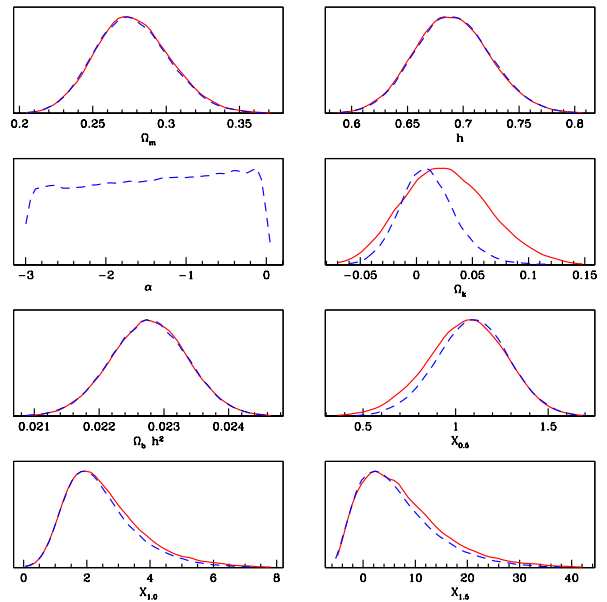,width=3.5in}\\
\caption[2]{\label{fig:Xz3_pdf}\footnotesize%
One dimensional marginalized pdfs of dark energy and cosmological
parameters from WMAP 5 year measurement of 
[$R(z_*)$, $l_a(z_*)$, $\Omega_b h^2$], 
182 SNe Ia, and the SDSS BAO scale measurement. 
Solid and dashed lines indicate $X(z)$ at $z>1.5$ given by 
$X(z)=X_{1.5}$ and $X(z)=X_{1.5}\,\exp[\alpha(z-1.5)]$ respectively.
}
\end{figure}

Table 2 shows the mean, rms variance, and correlation coefficients of 
($X_{0.5}$, $X_{1.0}$, $X_{1.5}$),
as well as the relative dark energy FoM$_r$ defined
in Eq.(\ref{eq:FoM_r}). 

\begin{table*}[htb]
\caption{Constraints on $X(z)$ parametrized by ($X_{0.5}$, $X_{1.0}$, $X_{1.5}$)}
\begin{center}
\begin{tabular}{lllllllllll}
\hline
$X(z>1.5)$ & $\mu(X_{0.5})$ & $\sigma(X_{0.5})$ & $\mu(X_{1.0})$ & $\sigma(X_{1.0})$ & 
$\mu(X_{1.5})$ & $\sigma(X_{1.5})$ & $\rho_{X_{0.5}X_{1.0}}$ & $\rho_{X_{0.5}X_{1.5}}$
& $\rho_{X_{1.0}X_{1.5}}$ & FoM$_r$ \\
$X_{1.5}$ &  1.059 &  0.213 &  2.556 &  1.215 & 7.503 &  8.037 &  -0.389 & -0.666 & 0.906 & 2.0771 \\
$X_{1.5}\,e^{\alpha(z-1.5)}$ &
1.091 & 0.195 & 2.436 &  1.121 & 6.533 &  7.351 & -0.303 & -0.609 & 0.895 & 2.402 \\
\hline
\end{tabular}
\end{center}
\end{table*}

\section{Summary and Discussion}

In order to compare current and future dark energy experiments
on the same footing, we have introduced a simple and straightforward 
definition for the Figure-of-Merit (FoM) of constraints on any set of 
dark energy parameters, Eq.(\ref{eq:FoM_r}), that is easily applicable 
to both real and simulated data. 

We recommend the adoption of two dark 
energy parametrizations in comparing different experiments: 
(1) A dark energy equation of state $w_X(z)$ linear in $a$, with its
values at $z=0$ and $z=0.5$, ($w_0$, $w_{0.5}$), as parameters
estimated from data [see Eq.(\ref{eq:w_0.5})].
We find that ($w_0$, $w_{0.5}$) are significantly less
correlated than ($w_0$, $w_a$) [see Table 1 and Fig.{\ref{fig:w0wc}}],
hence the factor of improvement in the FoM$_r$ [as defined in
Eq.(\ref{eq:FoM_r})] for ($w_0$, $w_{0.5}$) is a more reliable
indicator of the improvement in dark energy constraints than
the factor of improvement of FoM$_r$ for ($w_0$, $w_a$).
(2) The dark energy density function $X(z)=\rho_X(z)/\rho_X(0)$
parametrized by its values at $z=0.5$, 1.0, and 1.5,
($X_{0.5}$, $X_{1.0}$, $X_{1.5}$), for $z\leq 1.5$ [see Eq.(\ref{eq:Xz})],
and $X(z>1.5)=X_{1.5}$. We find that this flat cutoff
in $X(z)$ gives more conservative constraints on $X(z)$ than
parametrizing early dark energy with an extra parameter such
that cosmic curvature is constrained (see Fig.{\ref{fig:Xz3_pdf}}).

We have demonstrated the use of the FoM$_r$ [see Eq.(\ref{eq:FoM_r})]
for these two dark energy parametrizations 
[see Eq.(\ref{eq:w_0.5}) and Eq.(\ref{eq:Xz})] using
WMAP 5 year measurement of [$R(z_*)$, $l_a(z_*)$, $\Omega_b h^2$],
182 SNe Ia (from the HST/GOODS program, 
the first year Supernova Legacy Survey, and nearby SN Ia surveys), 
and the SDSS measurement of the baryon acoustic oscillation 
(BAO) scale [see Figs.{\ref{fig:w0wc}}-{\ref{fig:Xz3_pdf}}].
Dark energy is consistent with a cosmological constant at 68\% 
C.L. if one assumes the two-parameter dark energy equation 
of state model, $w_X(a)=3 w_{0.5}-2 w_0 + 3( w_0 - w_{0.5}) \,a$.
If one assumes dark energy density to be a free function
parametrized by its values at $z=0.5$, 1.0, and 1.5, then
dark energy deviates from a cosmological constant at $z =1.0$
at 68\% C.L., but is consistent with a cosmological
constant at 95\% C.L. (see Fig.{\ref{fig:Xz3_pdf}}).
This illustrates the importance of using the model-independent parametrization
in probing dark energy. Measuring $X(z)$ as a free function
from data allows us to detect epochs of variation
in dark energy density. It also allows us to constrain a broader
class of dark energy models than represented by $w_X(z)$;
for example, dark energy models in which $X(z)$ becomes negative in
the past or future, which are excluded by fiat if one only measures
$w_X(z)$ since $X(z)=\exp\{\int_0^z{\rm d}z' \, 3[1+w_X(z')]/(1+z')\}$
\cite{WangTegmark04}.
The two parameter dark energy equation of
state model (linear in $a$) implies strong assumptions about dark energy,
and is not sensitive to a transient variation in dark energy;
thus it is most useful in comparing forecasts
for future dark energy experiments under the simplest assumptions.

Future dark energy experiments from both ground and space
\cite{Wang00a,detf,ground,jedi,SPACE,Wang07}, together with CMB data 
from Planck \cite{planck}, will dramatically improve our ability 
to probe dark energy, and eventually shed light on the nature of dark energy.
Using both a linear dark energy equation of state [parameterized by
($w_0$, $w_{0.5}$)] and dark energy density function $X(z)$ as
a free function [parametrized by ($X_{0.5}$, $X_{1.0}$, $X_{1.5}$)]
provides a simple and balanced approach to exploring dark energy.
Proposed future dark energy experiments should be evaluated 
by comparing their FoM$_r$ for both ($w_0$, $w_{0.5}$)
and ($X_{0.5}$, $X_{1.0}$, $X_{1.5}$) to that of current data.

\bigskip

{\bf Acknowledgements}
I am grateful to Eiichiro Komatsu for sending me the covariance
matrix for [$R(z_*), l_a(z_*), z_*, \Omega_b h^2$] from
WMAP 5 year data, and 
for helpful discussions. I acknowledge the use of cosmomc
in processing the MCMC chains.

\end{document}